# Self-Modeling Based Diagnosis of Software-Defined Networks

José Manuel Sánchez, Imen Grida Ben Yahia
Orange Labs
Paris, France

Noel Crespi
Institut-Mines Télécom, Télécom SudParis, CNRS UMR5157
Evry, France

***Abstract*—Networks built using SDN (Software-Defined Networks) and NFV (Network Functions Virtualization) approaches are expected to face several challenges such as scalability, robustness and resiliency. In this paper, we propose a self-modeling based diagnosis to enable resilient networks in the context of SDN and NFV. We focus on solving two major problems: On the one hand, we lack today of a model or template that describes the managed elements in the context of SDN and NFV. On the other hand, the highly dynamic networks enabled by the softwarisation require the generation at runtime of a diagnosis model from which the root causes can be identified. In this paper, we propose finer granular templates that do not only model network nodes but also their sub-components for a more detailed diagnosis suitable in the SDN and NFV context. In addition, we specify and validate a self-modeling based diagnosis using Bayesian Networks. This approach differs from the state of the art in the discovery of network and service dependencies at run-time and the building of the diagnosis model of any SDN infrastructure using our templates.**

## I. INTRODUCTION

SDN (Software-Defined Networks) and NFV (Network Function Virtualization) is a novel phenomenon that is on the wish list of major industrial players (vendors, operators, content providers, software editors) as the means to achieve greater flexibility in managing the network, faster service deployment and provisioning while reducing operational costs. SDN is expected to pave the way towards network programmability by proposing network architecture based on abstraction, open interfaces, and control plane-data plane separation. On the other hand, NFV promise is to turn traditional network functions into virtual ones called Virtual Network Functions (VNF) and embed them into commoditized hardware. The combination of both approaches is commonly agreed to be the best solution and is the next industry move despite the preliminary stage of the management of such environment [2][3]. In particular, resiliency properties become fundamental for both technologies as discussed in [4]. We rely on self-diagnosis [5] as a first step towards resiliency. Self-diagnosis is an autonomic capability where the network is aware of any abnormal state and diagnoses itself to determine the root cause to perform the appropriate remediation or recovery actions. Those actions may be based on redundancy or diversity mechanisms, which bypass the presumed faulty network elements.

In the SDN and NFV context, the highly changing networks impose numerous diagnosis challenges, specifically how to detect continuous changes and update dependencies among virtual and physical resources.

The contribution of this paper is twofold: 1) definition of finer granularity templates that model the dependencies among SDN nodes (physical and logical) as well as smaller sub-components inside them (e.g. CPU, network cards, etc).

2) Specification and validation of a self-modeling approach that tracks changes on the network topology of the networked nodes and their corresponding VNFs at runtime.

The structure of the paper is as follows: Section II motivates our work and Section III summarizes the related work on model-based network diagnosis which includes the diagnosis techniques and the models that are in use. Section IV presents the proposed self-model based diagnosis framework including the template definition and the defined algorithm. Section V presents the experimental validation. Finally, Section VI summarizes our findings and outlines the future work.

## II. MOTIVATION

We consider an end-to-end service delivered to clients through SDN and NFV based networks. This combination of SDN and NFV raises the following questions with respect to diagnosis:

- The first challenge is how to model the dependencies among VNFs at runtime, as these dependencies may vary over time and depend on the service contracted by each user (Fig. 1 (a)).
- The second challenge is how to model at runtime the dependencies of the underlying network topology, when it is considered dynamic (Fig. 1 (b)). The network topology is dynamic due to several reasons like the connections and disconnections of users to the Access Points (AP), handovers, and especially VNF migrations and the orchestration of new services.
- The third challenge is how to model the two types of control in SDN, out-of-band and in-band. In out-of-band control,

the controller is directly connected to every switch by a dedicated control link, but in in-band control, the controller is only directly connected to the master switch, becoming this one an intermediate node that connects the controller with the rest of switches (slaves).

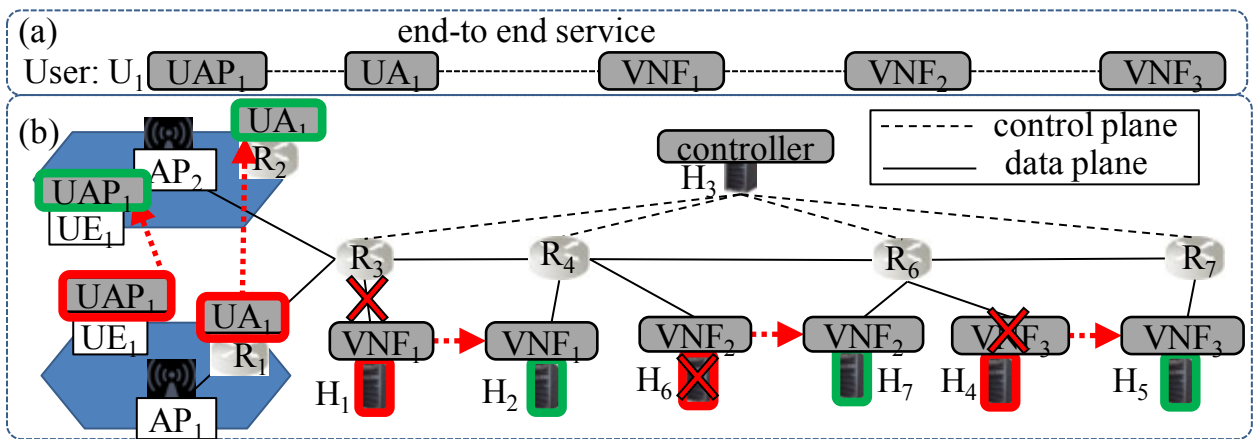

Fig. 1. SDN-NFV scenario: (a) end-to-end service, and (b) network topology

As it will be explained in section III, the most popular approach for diagnosis is the model-based diagnosis. Our approach targets to answer the challenges of diagnosis in the context of SDN and NFV and tracks the dynamic network topology. For this, we propose 1) a model that is suitable to handle the specific elements to monitor within a combined SDN and NFV scenario, and 2) a self-modeling algorithm to enable the automation of the diagnosis model.

## III. RELATED WORK ON MODEL-BASED DIAGNOSIS

The diagnosis model is generally a dependency graph. In general, the dependency graph is manually generated from an operational team's knowledge. This manual generation is valid for static network topologies, but not for dynamic and elastic networks such as those expected with SDN and NFV. A self-modeling approach is presented in the literature as the automatic generation of this model [6]. The dependency graph is first generated from a given data set (databases [7], genetic algorithms [8], or ontologies [9]). An inference engine (based on algorithms such as Bayesian Networks [6][10], Occam's Razor [12], or others) reasons then the dependency graph in order to extract the root cause.

In this paper, we focus on the generation at runtime of the dependency graph from the dynamic network topology. We consider the following works [6][10][11][12] which we explain and compare hereafter:

Hounkonnou et al. in [6] propose a self-modeling approach based on patterns to enhance Bayesian Networks (BN) and apply it to diagnose the IP Multimedia Subsystem (IMS). This approach generates offline a generic model (pattern) that is based on the four IMS layers (physical, functional, procedural and service). A pattern describes the dependencies among resources used by an IMS service. When a failure occurs in the IMS service, the algorithm locates online instances of that pattern in a given network topology and it generates the corresponding BN instance. However, this work assumes that the network topology remains static during the diagnosis process. Furthermore, the granularity of the diagnosis covers the four layers of IMS but it diagnoses only the network resource level, without considering smaller sub-components inside them. It is also worth mentioning that it does not consider virtual environments.

Bennacer et al. in [10] base their self-modeling approach on Chi-squared statistical tests. These tests learn the dependencies among modeled variables. They utilize a 'significance level' to decide their dependencies. This self-modeling approach considers physical symptoms as variables on each network node. However, the granularity of the diagnosis remains at the network node level, where smaller sub-components are not considered. Furthermore, the diagnosis is focused on the physical network nodes and is not considering the logical resources (e.g. virtual resources) running over them. In addition, inappropriate values of 'significance level' may lead to errors when building the dependency graph.

Kandula et. al. in [11] present a self-modeling approach based on templates. They define specific templates for each element such as a machine, an application process, a neighbor set and a path. Each template is characterized by several state variables to achieve a detailed diagnosis. Same as [6][10], the model granularity remains at the node level and does not consider smaller sub-components inside them.

Bahl et. al. in [12] models the dependencies among different services, but also the dependencies among network nodes given by the network topology. However, they do not consider smaller sub-components inside nodes or virtual resources.

This article advances the state of the art by describing a self-modeling based diagnosis to discover at runtime the dependency model of SDN-NFV infrastructures. It then models automatically an SDN based end-to-end service, its underlying network topology, and the type of control implemented (in-band or out-of-band). Compared to [6], which assumes that the network topology, services and configuration remain static, our proposal, in the context of SDN-NFV, assumes a continuous changing network topology. We then create the dependency graph by instantiating templates at runtime. Unlike [10], our self-modeling approach builds the dependency graph from the network topology, instead of statistical tests to avoid inaccuracies in the model. The idea of using different templates to describe different network elements was inspired by [11]. Our templates are tailored to suit the SDN elements' particularities and VNFs deployed in the network nodes. Contrary to [6][10][11][12], we consider a finer dependency model granularity, that considers smaller sub-components within a network node and their internal dependencies (physical and logical) to be able to diagnose a combined SDN and NFV environment where virtual and physical parts such as network cards and CPU need to be diagnosed. We use Bayesian Networks approach [13] for the root cause calculation which is enhanced with our self-modeling approach.

## IV. SELF-MODELING BASED DIAGNOSIS FRAMEWORK

We present a self-modeling based diagnosis framework for SDN (Fig. 2). We propose: (1) a template to model SDN

elements and (2) a self-modeling approach that builds automatically the model from the network topology.

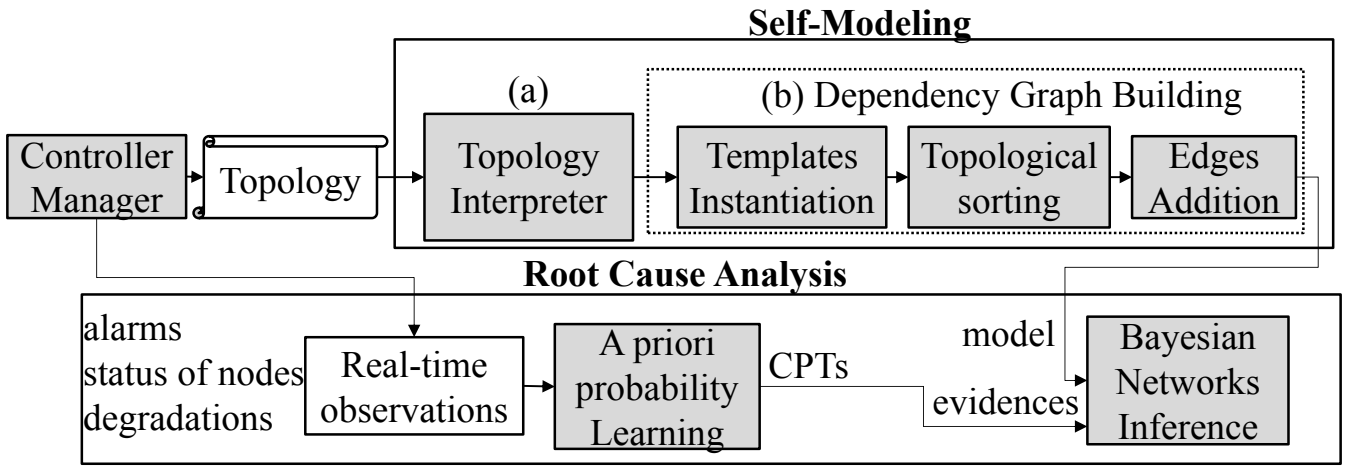

Fig. 2. Self-Modeling based diagnosis framework

The self-modeling building block is composed of the topology interpreter (a) and the dependency graph building (b) sub-blocks. This block automatically builds the dependency graph based on the network topology and the type of SDN control (in-band and out-of-band).

The root cause calculation building block finds the root cause through BNs. It receives a service alarm about service degradations or unavailability and correlates this alarm with network observations (the state of the physical and logical resources of each network node).

### *A. Background on Bayesian Networks*

BNs utilize probabilistic properties to perform reasoning in uncertain domains [13]. The model, the dependency graph, is represented by a set of vertices $V$ describing events interlinked by edges that represent the dependencies among vertices. The pair $B =< G, \theta >$ describes a BN. $G$ is the dependency graph, and $\theta$ contains the parameters of the BN, which take shape of Conditional Probability Tables (CPT) that specify the probability of every child vertex of the dependency graph given all value combinations of its parents. The prior probability of failure (p) is different for each vertex. The dependency graph $G$ is composed of observable and non-observable vertices. To reason over the dependency graph, we set the network observations into its observable vertices as evidence and the BN algorithm determines the most likely root cause(s).

### *B. Templates for modeling SDN networked elements*

We define a network element as any type of network nodes and links. We propose a template for each network element, so a template for network nodes and another for links. These templates describe the relationships between virtual and physical sub-components inside each network element.

#### *1) Network node template*

The template of a network node ($TN_n$ in Fig. 3(a)) is composed of a physical layer and a logical layer, following the TMF Information framework specifications [14].

- The physical layer encompasses the state of physical resources such as CPU and network cards. We consider two states for those physical sub-components inside network nodes: up or down.
- The logical layer encompasses the state of the VNFs running inside each node. We consider that network nodes perform one or several VNFs. For instance, the controller runs the appropriate network function that installs the rules in the switches as a VNF. A VNF relies on a software process, with suitable configuration settings. The logical layer contains three sub-layers in accordance with a three-state life-cycle for each VNF: initiated, configured, and activated. Initiated: where the underlying software process of the network function is launched; configured: where the VNF is set with the optimal attributes to perform the network function; and activated: where the network function is ready to accept any request.

The layers of the templates are manually predefined, but the number of network cards or running VNF per network node are retrieved from the network topology.

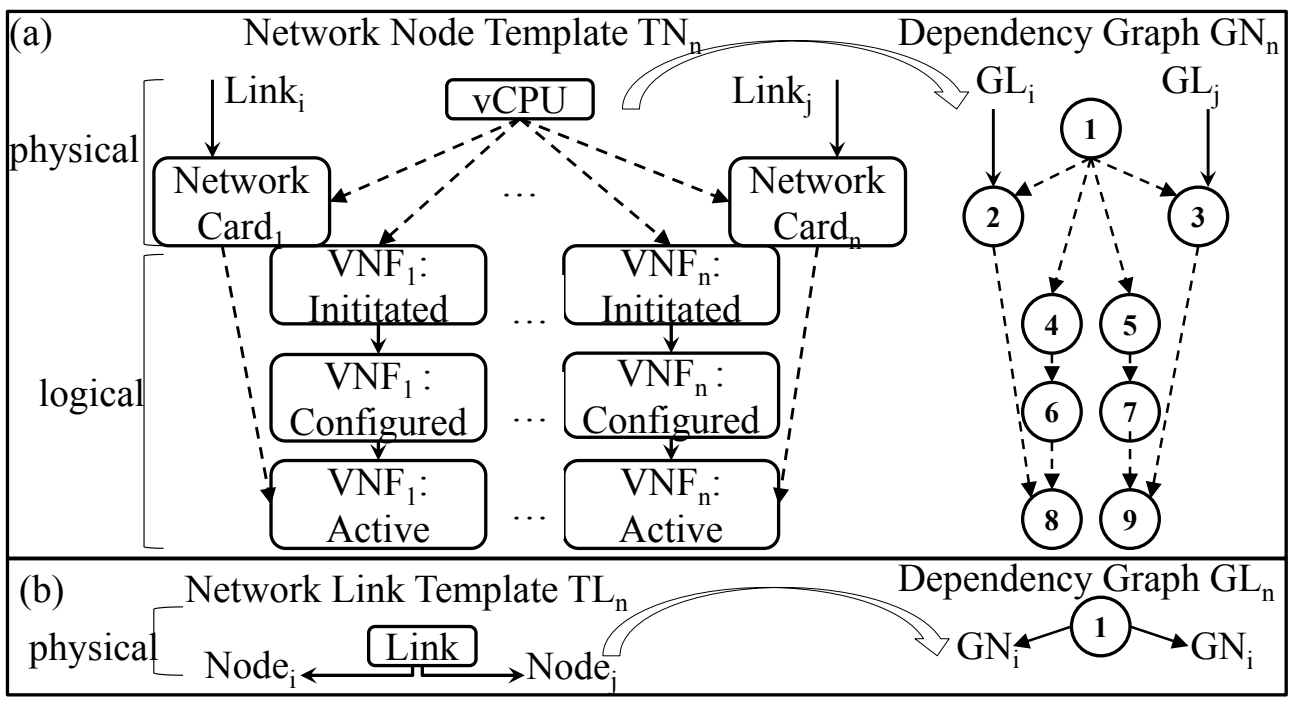

Fig. 3. Templates and Dependency Graphs: (a) network node and (b) link

Each type of network node discovered in the network topology (controllers ($C_1$… $C_n$), slave switches ($SS_1$… $SS_n$), master switches ($MS_1$… $MS_n$), hosts ($H_1$… $H_n$)) is an instance of this template.

#### *2) Link template*

The template of a network link ($TL_n$ Fig. 3 (b)) is simpler and it is composed of the physical layer and one vertex. Each type of link (control links ($CL_1$… $CL_n$), access links ($AL_1$… $AL_n$), and inter switch links ($IL_1$… $IL_n$)) is an instance of this template.

### *C. Self-modeling building block*

The self-modeling building block automatically models out-of-band and in-band SDN networks of any topology, by parsing the input data that contains the network topology, and automating the model creation process.

*1) Topology Interpreter:* The topology interpreter is a northbound application that requests the network topology from the controller. The controller answers this request by providing the network topology in a JSON format. The topology data structure depends on the controller: for example, OpenDaylight and a Floodlight controller provide two different data structures, differing in the number and type of fields and the field names (Fig. 4).

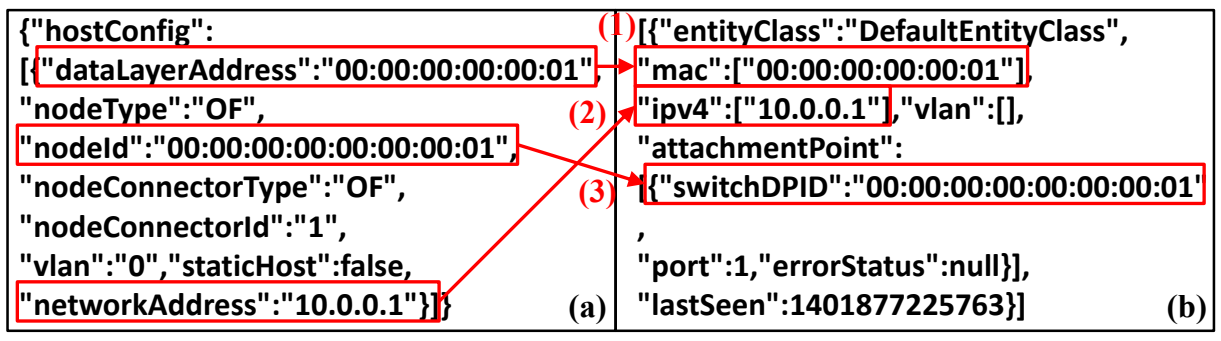

Fig. 4. JSON data structures provided by: (a) OpenDaylight, (b) Floodlight

Thus, in a second stage, the topology interpreter classifies each network element into one of the following types (controllers ($C_j$), slave switches ($SS_j$), master switches ($MS_j$), hosts ($H_j$), control links ($CL_j$), access links ($AL_j$), and inter switch links ($IL_j$). The result is then the network descriptor, which contains the network elements (nodes and links) at instant t, and the link descriptor, which specifies the end points of each link.

*2) Dependency graph building block:*
This block builds the global dependency graph (G) from the network descriptor. It is based on a three-step algorithm which we call: Template instantiation, topological sorting, and edge addition.

I) The template instantiation algorithm
This algorithm receives as input the network descriptor. It provides as output the global dependency graph *G* composed of the dependency graphs ($GL_n$ or $GN_n$) built from the templates of each network element. It follows this methodology for each network element found in the network descriptor:

- Identifies the type of network element
- Instantiates its corresponding template (TLn for links and TNn for nodes)
- Builds its associated dependency graph (GLn for links and GNn for nodes) from the instantiated template
- Appends the dependency graph into the global dependency graph G.

Each dependency graph ($GL_n$ or $GN_n$) contains edges called here $E_{INSIDE}$. $E_{INSIDE}$ edges (in dash in Fig. 3) depict the dependencies among sub-components inside each dependency graph ($GL_n$ or $GN_n$). Each vertex in $GL_n$ or $GN_n$ is a state variable described by a different CPT. A probability of failure ($P_{FAILURE}$) is defined for each vertex even though the parents vertices are functioning. As an example, $P_{FAILURE}$ for the network card (NC) vertex, P(NC=down) is p, despite the proper functioning of its parent vertex (vCPU).
The vertices of the global dependency graph G are not topologically sorted yet. We call these vertices as $V_{UNSORTED}$.

**Algorithm: Templates instantiation**
**IN**: Network Descriptor
**IN:** Templates
$\{T_{HOST}, T_{SLAVE}, T_{MASTER}, T_{CONTROLLER}, T_{ACCESS_LINK}, T_{CORE_LINK}, T_{CONTROL_LINK}\}$
**OUT**: Global Dependency Graph $G(V_{UNSORTED}, E_{INSIDE})$
**for** each element in the network descriptor
  inspection of type of element
  **if** element is of type link
    $TL_n \leftarrow$ instatiation of link template $\{T_{ACCESS_LINK}, T_{CORE_LINK}, T_{CONTROL_LINK}\}$
    $GL_n \leftarrow$ extract dependency graph of template $TL_n$
    $G \leftarrow$ append $GL_n$ to global dependency graph
  **else**
    $TN_n \leftarrow$ instatiation of node template $\{T_{HOST}, T_{SLAVE}, T_{MASTER}, T_{CONTROLLER}\}$
    $GN_n \leftarrow$ extract dependency graph of template $TN_n$
    $G \leftarrow$ append $GN_n$ to global dependency graph
  **end if**
**end for**

II) The topological sorting algorithm
It sorts topologically the vertices of the global dependency graph G. It receives as input the dependency graph G with non-topologically ordered vertices ($V_{UNSORTED}$) and it provides as output the same global dependency graph but topologically sorted ($V_{SORTED}$). As an example, $GN_n$ and $GL_n$ (Fig. 3) present a topological order when separated, because any vertex index is repeated and parents are numbered before children vertices. However, when both are combined to obtain the global dependency graph G, the topological order is not respected, as both $GN_n$ and $GL_n$ contain repeated vertex indexes (e.g. both contain value '1'). The topological sorting algorithm solves this issue.

**Algorithm: Topological Sorting**
**IN:** Global Dependency Graph $G(V_{UNSORTED}, E_{INTRA})$
**OUT:** Global Dependency Graph $G(V_{SORTED}, E_{INTRA})$
**for** each dependency graph appended to G
  **for** each layer in template
    obtain vertices of appended graph at current layer
    sort its vertices topologically
  **end for**
**end for**

III) The edge addition algorithm
This algorithm adds the dependencies between the previously appended dependency graphs ($GN_n$ and $GL_n$) by the template instantiation algorithm. It receives as input the link descriptor and the global dependency graph (topologically sorted).

- It puts one dependency graph of link GLn in between the two dependency graphs of the nodes GNn to be connected (these nodes are given in the link descriptor).
- It then adds two $E_{INTER}$ edges from the dependency graph of the link $GL_n$ to the two dependency graph of nodes $GN_n$.

$E_{INTER}$ depicts the dependencies between $GL_n$ and $GN_n$. In Fig. 3(a), the dependency graph of one node $GN_n$ has two incoming $E_{INTER}$ edges from the dependency graphs of links $GL_i$ and $GL_j$. In Fig. 3(b), the dependency graph of one link $GL_n$ has two outgoing edges $E_{INTER}$ towards the dependency nodes of $GN_i$ and $GN_j$.

**Algorithm: Edge addition**
**IN:** Link Descriptor, Global Dependency Graph $G(V_{SORTED}, E_{INSIDE})$
**OUT:** Global Dependency Graph $G(V_{SORTED}, E_{INSIDE}, E_{INTER})$
**for** each link in Link Descriptor
  extract end points attached to link
  **for** each end point in link
    $E_{INTER} \leftarrow$ add edge from $GL_n$[link,link] to $GN_n$[node,card]
  **end for**
**end for**

## V. VALIDATION OF DIAGNOSIS MODULE

We test our self-modeling based diagnosis in a centralized SDN architecture based on a Floodlight controller. This module runs over the controller for two reasons: (1) to have a global view of the network, and (2) to keep the diagnosis framework independent from any specific southbound protocol. The network topology is obtained via the northbound interface (REST API) trough passive monitoring, to avoid introducing traffic overhead like ping tool. We use Mininet to simulate the SDN network.
First, we prove that our self-modeling algorithm can interpret both the topology and the control type of SDN (out-of-band and in-band). Next, we study the scalability of this algorithm and finally we validate the diagnosis results and their variation under changing network conditions.

## *A. Self-modeling Validation*

We test the model generation of a linear topology with two switches and a controller with two hosts connected under out-of-band (Fig. 5 (a)) and in-band control (Fig. 5 (b)).

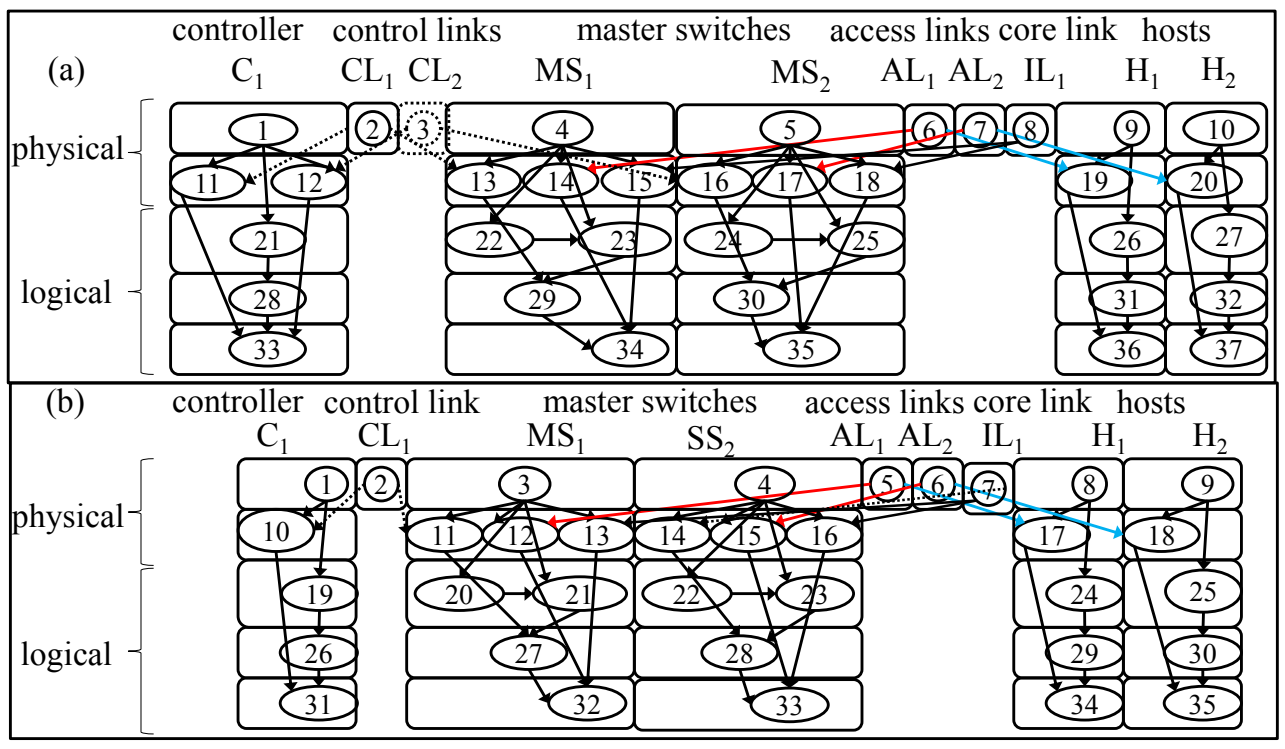

Fig. 5. Dependency graphs of linear topology in: (a) out-of-band control, (b) in-band

Fig. 5 shows the resulting global dependency graph G built by the self-modeling algorithm. This graph G is topologically ordered. We analyze the output of the algorithm hereafter:

-The self-modeling algorithm interprets the type of control:

In out-of-band control, the self-modeling algorithm instantiates two control links (instances: $CL_1$, $CL_2$, vertices: 2, 3), where $CL_1$ connects both the network card of the master switch (instance: $MS_1$, vertex: 13) and the network card of the controller (instance: $C_1$, vertex: 11). The controller template ($C_1$) has two network cards connected to two switches (vertices: 11, 12) through the control link instances $CL_1$ and $CL_2$. In in-band control, it only instances one control link (instance: $CL_1$, vertex: 2) because the controller is only connected to the master switch ($MS_1$). The controller template then has one network card (vertex: 10), which is connected to the network card of the master switch instance (vertex: 11) through the control link instance. The other switch is slave (instance: $SS_1$) and communicates to the controller trough the link $IL_1$ that connects to the master switch.

-The self-modeling algorithm interprets the network topology: For both types for control, it connects both switches $MS_1$ and $MS_2$ through the inter switch link instance ($IL_1$). It connects both hosts' instances ($H_1$ and $H_2$) to their respective access links ($AL_1$ and $AL_2$).The algorithm can automatically generate ring, star, linear and tree network topologies for different numbers of hosts and switches.

-Scalability:

We study the growth in number of vertices (*V*) of the global dependency graph *G* for linear and tree topologies for out-of-band control. We analyze both topologies for a varying number of connected hosts ($N_{HOSTS}$) from 4 up to 256. The number of network elements ($N_{ELEMENTS}$) (nodes and links) is the same for both topologies $N_{ELEMENTS}=3N_{SWITCHES}+2N_{HOSTS}$. The number of vertices in the global dependency graph G is: $V=V_{CONTROLLER}+V_{SWITCHES}N_{SWITCHES}+V_{HOSTS}N_{HOSTS}+V_{LINK}(2N_{SWITCHES}+N_{HOSTS}-1)$. If we particularize with the values for the aforementioned topology (Fig. 5 (a)) in out-of-band control: 5 vertices per host template ($V_{HOST}=5$), 8 vertices per switch template ($V_{SWITCHES}=8$), 1 vertex per link template ($V_{LINK}=1$) and 5 vertices per controller template ($V_{CONTROLLER}=5$), this equation becomes $V=5+10N_{SWITCHES}+6N_{HOSTS}$, which explains the linear trend of vertices with the number of hosts described in Table II.

TABLE I. NUMBER OF VERTICES (V) AS A FUNCTION OF THE NUMBER OF HOSTS ($N_H$)

| **Topology/$N_H$** | **4** | **8** | **16** | **32** | **64** | **128** | **256** |
|---|---|---|---|---|---|---|---|
| Tree | 62 | 130 | 266 | 538 | 1082 | 2170 | 4346 |
| Linear | 72 | 140 | 276 | 548 | 1036 | 2180 | 4356 |

We study the speed of the self-modeling algorithm as a function of the number of network elements ($N_{ELEMENTS}$). We launched the self-modeling algorithm for both linear and tree topologies in out-of-band control, ranging from 15 up to 500 network elements. We averaged the computing time 20 times per topology to obtain more reliable figures. Fig. 6 shows an exponential trend in the growth of self-modeling time with the number of elements for both topologies. Linear topologies scale better than tree topologies, but in both cases the self-modeling time remains less than 30 seconds.

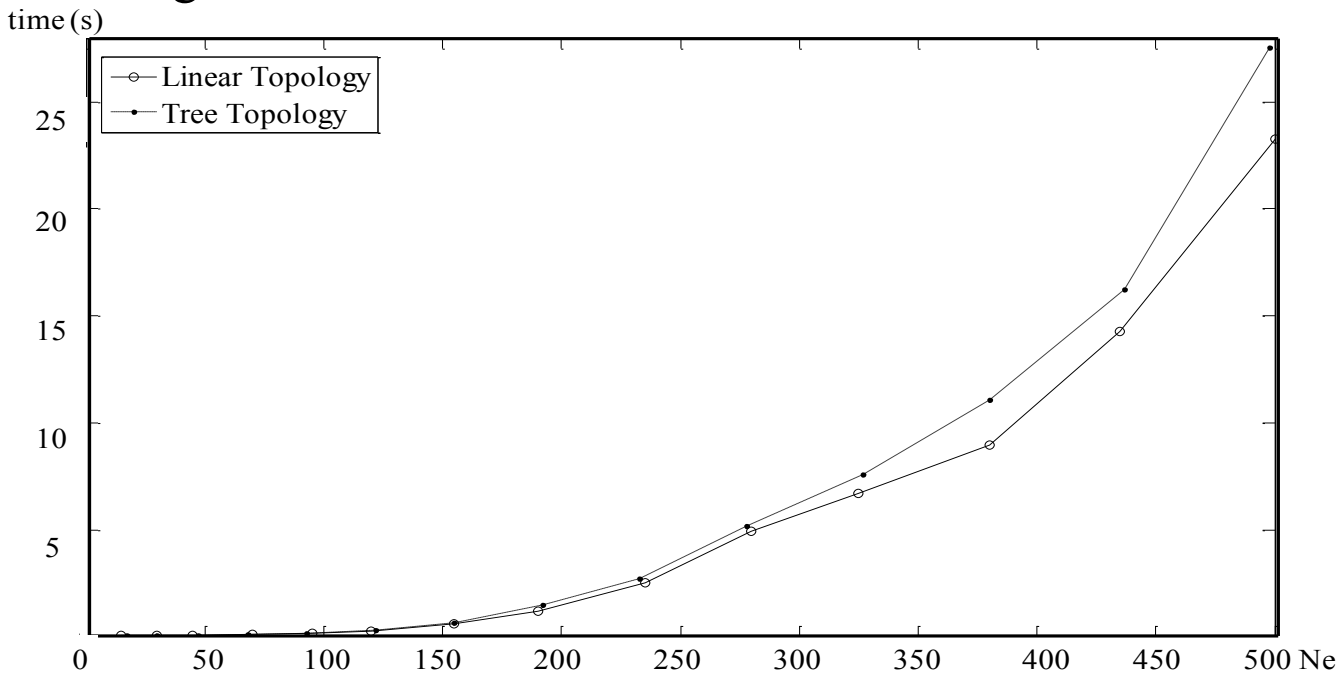

Fig. 6. Speed as a function of the number of elements

## *B. Validation of root cause analysis*

We validate the diagnosis module in a linear topology (Fig. 7) in out-of-band control in two different scenarios.

In the first scenario, we study two cases, where we force certain failures on the SDN infrastructure and test if the BN engine can pinpoint accurately the failed element. It receives an alarm about this failure on SDN infrastructure and it is fed with the states of network cards of all the network nodes as network observations. The BN engine correlates these observations with this alarm to find the root cause.

In case 1 (Fig. 7(a)), the actual root cause is a total shutdown of the controller. The BN engine determines that the most probable root cause is the controller (94.2 %). It determines that the CPU (31.4 %), the VNF (31.4 %) or its configuration (31.4 %) could be the source of the failure. In case 2 (Fig. 7(b)), the actual root causes are three simultaneous link failures in the control link $CL_1$ and access links $AL_1$ and $AL_2$. The BN engine pinpoints those affected links $CL_1$, $AL_1$, $AL_2$ as the most probable root causes (31.1 %), having discarded the rest of elements.

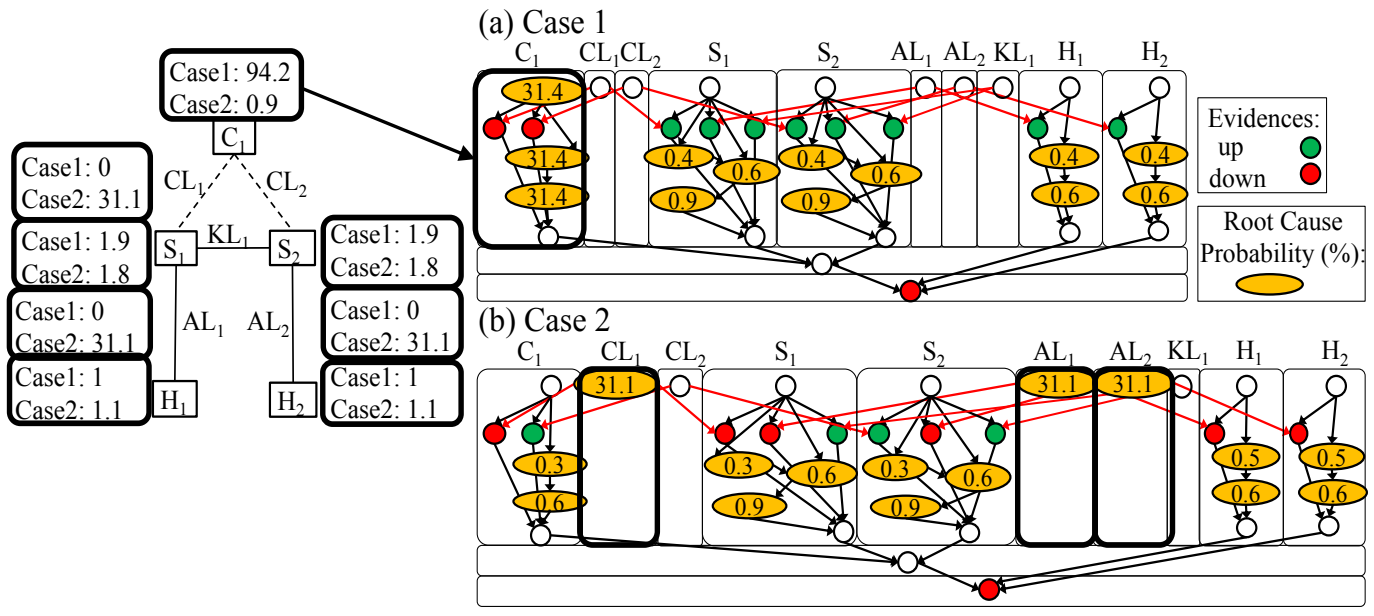

Fig. 7. Root Cause analysis: (a) Case 1, (b) Case 2

In the second scenario, we consider degradations in the SDN infrastructure affecting a service between $H_1$ and $H_2$. This degradation may be explained by monitoring the CPU use on the network nodes, so the BN engine incorporates these observations. As consequence, the calculated root changes as a result of changes on these observations. We run the BN engine with two different CPU conditions on the nodes, shown in Fig. 8: (a) a non-loaded controller (CPU use 5%) with host $H_2$ heavily-loaded (CPU use 95%) and the rest of nodes with a CPU use between 5% and 95%), and (b) a heavily-loaded controller (CPU use 95%) and the rest of nodes with a CPU use between 5% and 95%. In situation (a), the BN engine determines that host $H_2$ is the most probable root cause (96.6%) due to its high CPU use, and so it discards all the links (1 %) as well as the controller (7.9 %) as probable causes. In the heavy load of CPU use in the controller (b), the BN engine selects the controller as the most probable root cause to explain the degradation on the SDN infrastructure (a transition of root cause probability from 7.9 % to 96.9 %).

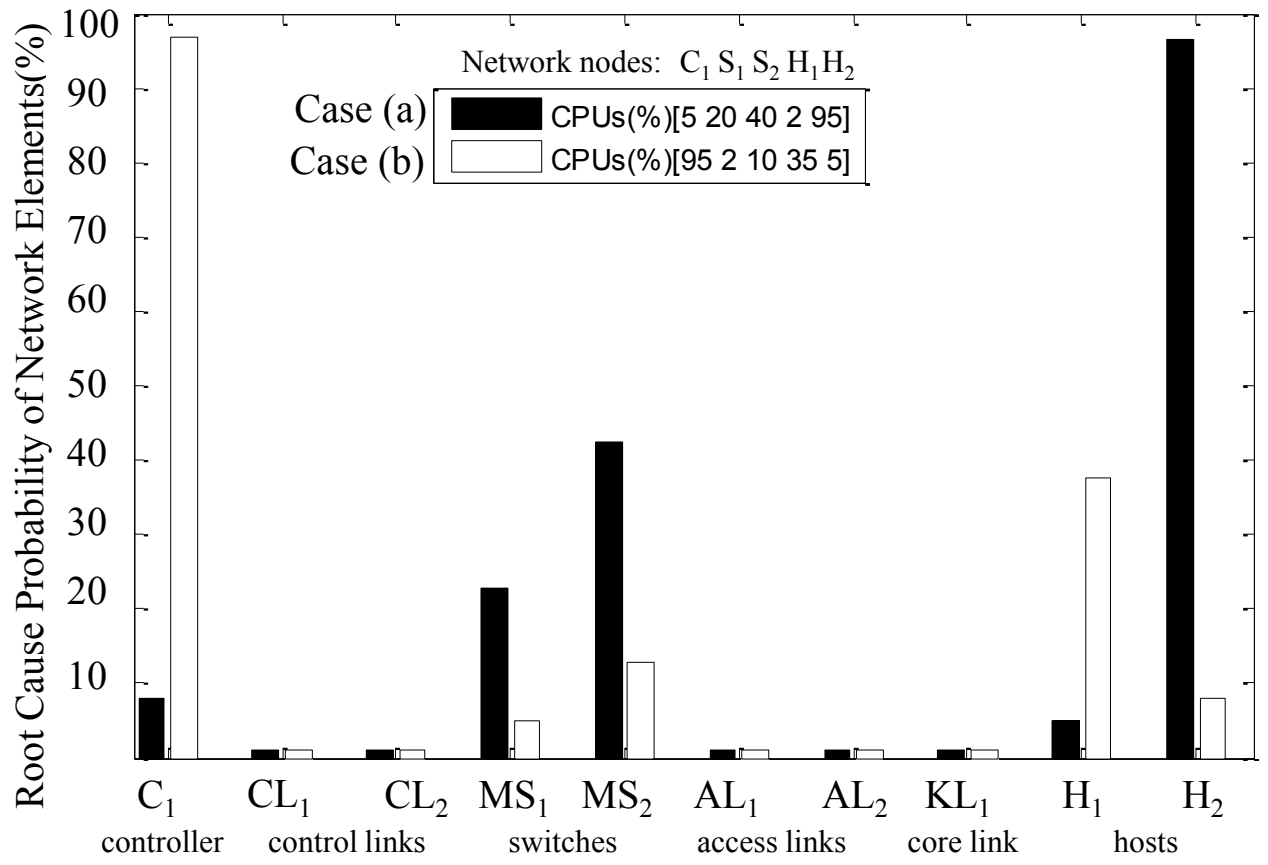

Fig. 8. Root Cause Analysis on changing CPU conditions

## VI. Conclusions and Future Work

This paper considers solving two major problems towards self-diagnosis and resilient networks in the context of SDN and NFV: in such context it is needed to define a template or a model that describes the managed elements including physical, virtual infrastructure and other inner details such network cards, or CPU. To fill this gap, we define a template with finer granularity describing the essential managed elements within SDN and NFV. Furthermore, we specify and validate a self-modeling diagnosis that builds automatically at runtime the diagnosis model (dependency graph), which answers the challenges of updating the diagnosis model to identify the root causes. Our approach is suitable to any network topology and to any control type in SDN. In addition, it is independent from the controller implementation (e.g. Floodlight or OpenDaylight).

Regarding future work, we will focus on the following points:

- Reducing the uncertainty of the diagnosis: As a result of the finer granularity level of our proposed model, the uncertainty of the diagnosis is high. We foresee to adapt the methodology of Hounkonnou et. al. [6] to reduce the uncertainty by updating the model progressively with observations obtained from new clients.
- Learning network element templates to diagnose new root causes: we target to add learning mechanism to our self-modeling based diagnosis. Statistical tests can be used as discussed in [10] to learn automatically node templates allowing the diagnosis of unexpected new root causes.
- Modeling VNFs dependencies: Extension of the self-modeling algorithm to model the VNF forwarding graphs that compose the service of each client online.

## I. Acknowledgments

This work is partly funded by the French ANR under the ANR-14-CE28-0019 REFLEXION project